\documentclass[preprint2]{aastex}

\def\del#1{}      

\usepackage{epsf}

\begin{document}

\title {3C 129 at 90cm: Evidence for a Radio Relic?}

\author{W.M. Lane, N.E. Kassim}

\affil{Naval Research Lab, Code 7213, 4555 Overlook Ave. SW, Washington, DC, 20375} 

\email{lane@rsd.nrl.navy.mil, kassim@rsd.nrl.navy.mil}

\author{T.A. En{\ss}lin}

\affil{Max-Planck-Institut f{\"u}r Astrophysik,
Karl-Schwarzschild-Str.1, 85740 Garching, Germany}

\email{ensslin@MPA-Garching.MPG.DE}

\author{D.E. Harris}

\affil{Harvard-Smithsonian Center for Astrophysics, 60 Garden Street,
Cambridge, MA 02138}

\email{harris@head-cfa.harvard.edu}

\and

\author{R.A. Perley}

\affil{National Radio Astronomy Observatory, Socorro, NM 87801, USA}

\email{rperley@nrao.edu}

\begin{abstract}

We present a new wide-field map of the radio galaxy 3C 129 and its
companion galaxy 3C 129.1 at $\lambda = 90$ cm.  We see a distinct
steep-spectrum feature near the head of 3C 129, extending in a
direction perpendicular to the radio tails.  We propose that this
Crosspiece might consist of fossil radio plasma, which has been
re-energized by the compression of the bow shock of the supersonically
moving galaxy 3C 129. One possible origin of the fossil radio plasma
could be the tail of a nearby head-tail radio galaxy. We discuss the
implications of, and give testable predictions for, this scenario.
\end{abstract}


\newpage

\section{Background}

The radio galaxy 3C 129 and its companion 3C 129.1 are part of an
X-ray galaxy cluster at z=0.021 (4U 0446+44).  The cluster lies in the
galactic plane (b$=0.5^{o}$, l$=160^{o}$) and is only known to include
one other member galaxy (WEIN 048) which lies to the south of 3C 129
\citep{nil00}.  Because of their latitude, the cluster and the radio
sources have been excluded from many optical studies.

Both radio sources have jets extending beyond their elliptical host
galaxies and into the intracluster medium (ICM).  3C 129 shows
prototypical narrow-angle-tail (NAT) morphology, with a plume-like
double-tail extending nearly 30' at $\lambda \sim 90$ cm.  It has a
total flux density of $\sim 5.3$ Jy at 1400 MHz \citep{wb92}.  3C
129.1, although considerably smaller, is a wide-angle-tail (WAT)
source, with a total flux density of $\sim 1.9$ Jy at 1400 MHz
\citep{condon98}.

ROSAT X-ray data place the cluster center just to the southwest of 3C
129.1.  The X-ray contours are slightly elongated in the East-West
direction, suggesting that the cluster may have undergone recent
merger activity.  In addition, the X-ray contours are distorted near
the brightest part of 3C 129; Leahy and Yin (2000) suggest that this
may be caused by the radio plasma moving through the ambient cluster
gas at speeds greater than the local speed of sound. The large-scale
morphology of 3C 129 bends in a manner which is inconsistent with a
naive tracing of its trajectory in the cluster potential. Cowie and
McKee (1975) propose that buoyancy of the radio plasma in the ICM
atmosphere may have helped shape the radio tail. This would imply that
3C 129 can be used to test the mechanical properties of radio plasma.

\section{Data}

\begin{figure}[h]
\centering
\epsfxsize=1.0\columnwidth
\epsfbox{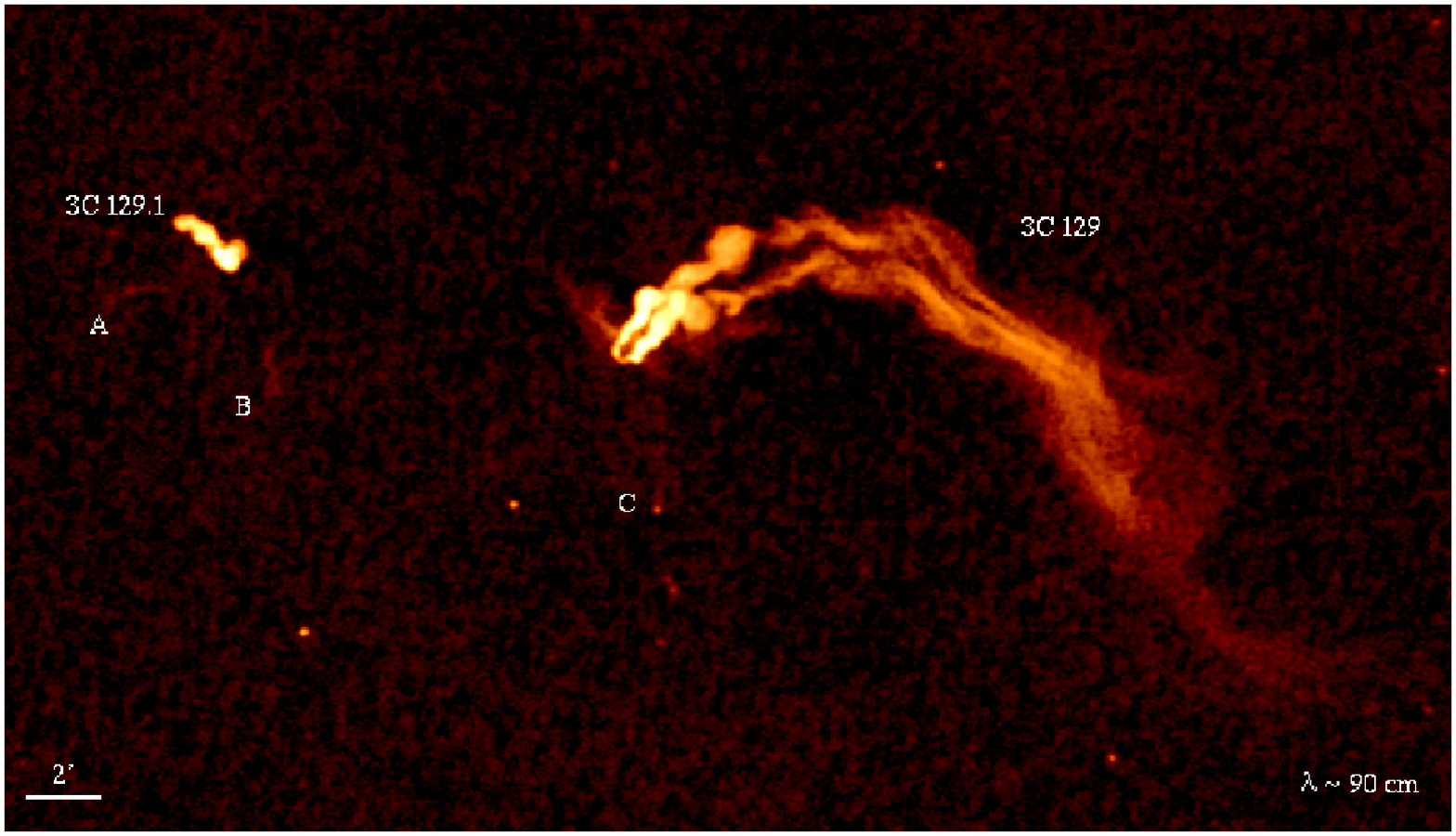}
\caption {\small A wide-field image of 3C 129 and its companion 3C
129.1 at $\lambda \sim 90$ cm.  The faint objects labeled A,B, and C
have all been identified at other frequencies and are real.  Note the
small perpendicular extension to either side of 3C 129, which is the
Crosspiece discussed in the text.  \label{fig:3c129mark}}
\end{figure}

During the late 1990's we obtained roughly 175 ``snapshot''
observations of this source using the Very Large Array
(VLA)\footnote{operated by The National Radio Astronomy Observatory, a
facility of the National Science Foundation operated under cooperative
agreement by Associated Universities, Inc.}, each lasting a few
minutes.  The data at 330 MHz were intended to provide
phase-referencing information for simultaneous observations at 74 MHz.
Slightly more than half of the data were taken in A-array; the rest
are nearly equally divided between B- and C-arrays.  Standard
calibration observations were used, and the data were mapped using
wide-field imaging techniques in AIPS.  The resulting image, shown in
Figure ~\ref{fig:3c129mark} has nearly complete UV-coverage and
roughly 8'' resolution at a frequency of 330 MHz.  The rms noise
in the map is $\sim 0.75$ mJy, and the signal-to-noise ratio
is S/N$\sim 400$

\section{Sources A, B, and C}

The two faint sources marked A and B also appear in low resolution 74
MHz maps of this source (K. Blundell 2002, in prep.).  Comparison with
the 74 MHz image gives a spectral index $\alpha \approx 2.4$ for both
sources ($S \propto \nu^{-\alpha}$).  Neither source appears in the
NVSS catalogue \citep{condon98}.  The 2.5 mJy detection limit of that
catalogue implies spectral indices $\alpha \geq 1.7, 2.1$ for sources
A and B respectively, in agreement with the 74 MHz index.  We suggest
that these faint sources are two buoyant radio lobes produced by
earlier activity of 3C 129.1. Their comparable sizes, luminosities,
and similar distances to 3C 129.1 support this conclusion. Future
measurements of their high frequency radio spectral cutoffs would
provide a date for the corresponding phase of 3C 129.1, and provide an
estimate of the buoyant rise velocity of these old lobes.

Source C is too faint to be detected in the existing 74 MHz data.  It
is identified in the NVSS catalogue as a 3.2 mJy point source.  In our
330 MHz map it is slightly extended, and has an integrated flux of
27.8 mJy.  Combining these two fluxes we calculate a spectral index
$\alpha = 1.5$.  This source is coincident with the optical galaxy
WEIN 047, and is also an IRAS source.

\section{The Crosspiece}

\begin{figure}[t!]
\centering
\epsfxsize=1.0\columnwidth
\epsfbox{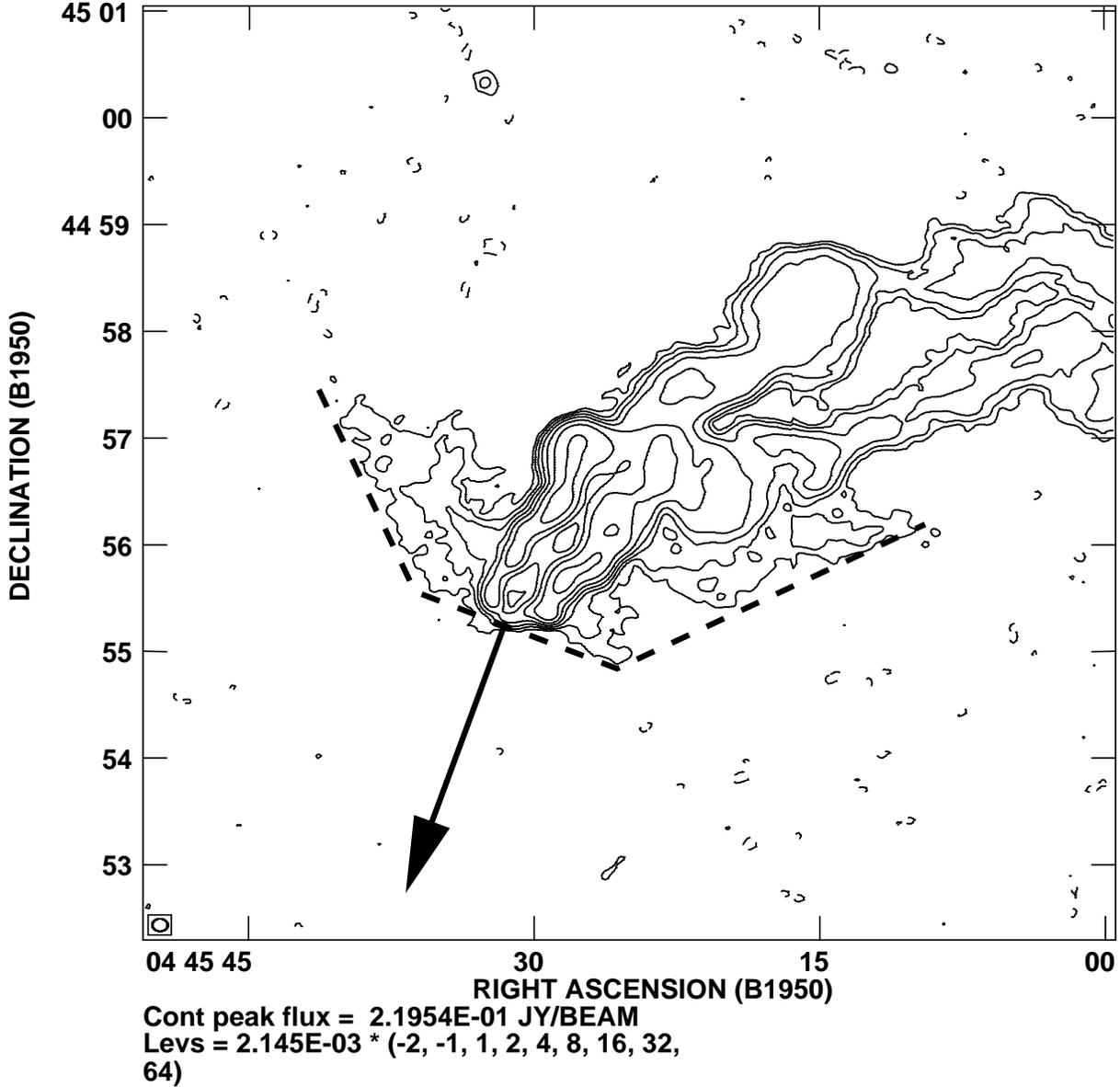}
\caption{\small A closer look at the 90cm map near the head of 3C 129,
showing the Crosspiece feature extending to both sides.  Contour
levels are multiples of the 3$\sigma$ noise in the map. The dashed
line sketches the suggested location of the shock wave, which would
have a Mach cone opening angle of $\theta_{\rm Mach} \approx
90^{\circ}$. If this is correct, the arrow indicates the motion of 3C
129 within the plane of the sky. As argued in the text, the
3-dimensional motion should not differ more than $\sim 10^\circ$ from
this. Note, that the radio trails of 3C 129 are likely buoyant in the
cluster ICM, so that their orientation need not be exactly parallel to
the galaxy motion. This is supported by the strong bending of the
large scale morphology of the trails \citep{cm75}.
\label{fig:3c129Fish}}
\end{figure}

Figure ~\ref{fig:3c129Fish} shows a contour image of the region near
the head of 3C 129.  There is a very clear feature just behind the
head of the radio source, and running perpendicular to the tails.
Its nature is still uncertain, but it appears to extend over
both tails, so we will refer to this as the Crosspiece.

The contours in figure ~\ref{fig:3c129Fish} are at multiples of the
3$\sigma$ noise level in the map itself: it is clear both that the
Crosspiece is highly significant and that there are no other
suspiciously deep holes or bright regions in the area.  Further
confirmation of the reality of this feature was found in a paper by
J{\"{a}}gers \& de Grijp (1983) who made an early map of this source
using the Westerbork Synthesis Radio Telescope at a wavelength of
$\lambda \sim 50$ cm.  They identify a small projection on one side of
3C 129 which corresponds in position to the Crosspiece.  Using a
zeroth order beam size correction, we estimate a spectral index
between 330 MHz and 600 MHz of $\alpha \approx 3$ ($S \propto
\nu^{-\alpha}$).

Although the estimate is imprecise, it is clear that this feature is
very steep spectrum.  Few other radio maps of this source exist at
intermediate frequencies, but there are some high quality maps at
$\lambda \sim 6$ cm \citep{tay01}.  Not unexpectedly if the feature is
truly steep spectrum, these show no hint of the Crosspiece.  The only
existing map at 1400 MHz \citep{j87} is of very poor quality; we have
recently received time to make a new one (Harris et al. 2002, in
prep.).

\subsection{A Scenario for the Crosspiece's Genesis}

What then is the nature of the Crosspiece?  It is always possible that
it is simply a background radio source which happens to lie along the
same line of sight as 3C 129.  Given its steep spectrum nature, it
would most likely be a high-redshift radio galaxy or possibly even a
cluster.  However its large angular extension (3'-4') makes this seem
somewhat unlikely.  The location and orientation of the Crosspiece
near the head of 3C 129 suggests a physical connection between the
two. In the following we introduce a possible formation scenario and
discuss its implications.

The morphology, the low surface brightness, and the steep spectrum
clearly distinguish the Crosspiece from the usual radio plasma outflow
of 3C 129. It has more similarities with a so called {\it cluster
radio relic}. Cluster radio relics are steep spectrum radio sources,
typically located in peripheral regions of galaxy clusters, which need
not be associated with any parent galaxies (for a review of the
observations, see \cite{fg96}).  These radio relics are believed to
trace shock waves in the ICM and most likely consist of shock
compressed fossil radio plasma
\citep{1998A&A...332..395E,1999ApJ...518..603R,2001A&A...366...26E,kassim2001,eb2001}.

Remnant radio plasma which is not too old is able to revive its radio
emission after a shock compression
\citep{2001A&A...366...26E}. Because the fossil's radio plasma
internal sound speed is likely to be much higher than a typical
cluster shock speed, the compression is adiabatic. Nevertheless, the
compression factor can be high due to the soft equation of state of
relativistic plasma (Pressure $\sim$ density$^{\gamma}$ with a soft
$\gamma_{\rm rel.}  \approx 4/3$ instead of the nonrelativistic less
compressable $\gamma_{\rm non-rel.} = 5/3$), and the high energy
cooling cutoff of the still relativistic electron population can be
shifted adiabatically to observable energies. The simultaneous
amplification of the magnetic fields during the compression also
increases the characteristic synchrotron frequency of the highest
energy electrons. This relic formation scenario has recently been
supported by the morphological agreement between numerically simulated
and observed cluster radio relics \citep{slee2001,eb2001}.

\subsection{Theoretical Considerations}

Here we propose that the bow shock of 3C 129 itself might have revived
the radio emission of the Crosspiece.  In order to explain the bending
of 3C 129's jets, a velocity of 3C 129 with respect to the ICM of
$v_{\rm gal} = 500 - 3000$ km s$^{-1}$ is assumed by various authors
\citep{cm75,1995AJ....110...46D,ly00}. The ICM sound velocity is
$c_{\rm s} = 1217 \pm 22$ km s$^{-1}$ for the cluster temperature of
$kT_{\rm ICM} = {5.5\pm 0.2 \,\rm keV}$ \citep{ly00}, and a Helium
mass fraction of $\chi_{\rm He} = 0.25$, so 3C 129 could have a Mach
number as large as $M=v/c_{\rm s} \leq 2.5$.  3C 129's Mach cone
should therefore have a minimum opening angle of $\theta_{\rm Mach}
\geq 2 \arcsin(M^{-1}) \geq 50^\circ$.  This angle would increase if
3C 129's velocity is lower.

In order to heat a cold infalling plasma to the cluster temperature,
$kT_{\rm ICM}$, in an accretion shock wave, the typical infall
kinetic energy per particle has to be $E_{\rm inf} =
\frac{\mu}{2}\,v_{\rm inf} \equiv \frac{3}{2}\,kT_{\rm ICM}$, where
$\mu$ is the mean molecular mass of the gas. Therefore the infall
velocity of matter onto the cluster must be $v_{\rm inf} \approx
\sqrt{3\,kT_{\rm ICM}/\mu} = 1630$ km s$^{-1}$. This corresponds to a
(post-accretion shock) Mach number of $M = 1.34$.  3C 129 is a member
of an infalling galaxy cluster and could have an additional velocity
component due to the infalling cluster's internal velocity
dispersion. Nevertheless, we believe that a velocity close to the
typical infall velocity is more likely than $v_{\rm gal} > 2500$ km
s$^{-1}$.

3C 129's line-of-sight velocity $v_\|$ should be given by the
difference between its redshift ($z_{\rm 3C 129} = 0.020814$) and that
of 3C 129.1 ($z_{\rm 3C 129.1} = 0.022265$). The latter is located at
the cluster's X-ray center, has a nearly symmetrical radio appearance,
and is therefore very likely to be at rest with respect to the
cluster. Because the line-of-sight velocity for 3C 129 estimated in
this way, $v_\| = 435\,{\rm km\, s^{-1}}$, is low compared to the
typical infall velocity, the supersonic motion of 3C 129 should be
close to the plane of the sky ($19^\circ$ for Mach number $M = 1$,
$13^\circ$ for $M = 1.4$, and $7.5^\circ$ for $M=2.5$).  Therefore, if
the Mach cone can be assumed to be fully visible, projection effects
of the revived fossil radio plasma should not have significantly
altered its apparent (sky-projected) opening angle.

A visual inspection of the radio map (see Fig. \ref{fig:3c129Fish})
suggests an opening angle of $\theta_{\rm Mach} \approx 90^\circ$,
corresponding to $M = 1.4$ or $v_{\rm gal} = 1720$ km s$^{-1}$. We
note that that the underlying assumption of this estimate, the full
appearance of the Mach cone in the radio map, is not necessarily true,
because the radio morphology can be affected by the morphology of the
revived radio plasma. Nevertheless, the Crosspiece's morphology looks
suggestive, and the derived velocity seems reasonable because it is
close to the typical infall velocity of the cluster.

In order for the radio emission to be revived at a frequency of at
least $\nu = 1420$ MHz, where we see part of the Crosspiece (Harris
et al. 2002, in prep.), the radio plasma must be younger than
\begin{equation}
\label{eq:age}
t_{\rm max} = \frac{39 \, {\rm Myr} \,C^{2/3}}{( u_B + u_{\rm cmb})/({\rm
 eV\,cm^{-3}})}\, \sqrt{ \frac{B/{\rm \mu G}}{\nu/{\rm GHz}}} \,.
\end{equation}
\citep{2001ApJ...549L..39E}.  Otherwise the spectral cutoff is below
the observing frequency. The terms $u_{\rm cmb} = 0.26\,(1+z)^4\,{\rm
eV\,cm^{-3}}$ and $u_B= B^2/(8\,\pi) = 0.025 \,(B/\mu{\rm G})^2\,{\rm
eV\,cm^{-3}}$ denote the CMB energy density and the magnetic field
energy density before the radio plasma compression.  The compression
factor is called $C$, and can be estimated from pressure equilibrium
with the environment before and after the passage of the fossil radio
plasma through the shock wave. With a non-relativistic environment,
and an ultra-relativistic fossil radio plasma gas equation of state,
we find
\begin{equation}
C = \left( \frac{5 M^2 -1}{4} \right)^{\frac{3}{4}}\,,
\end{equation}
where we assume that the shock has the same Mach number as 3C 129.
This gives $C = 1.8$ for $M=1.4$, and $C= 4.6$ for $M = 2.5$.
Inserting this into Eq. \ref{eq:age}, we find that that the fossil
radio plasma has to be younger than $t_{\rm max} = 160$ Myr for $M=
1.4$ and $t_{\rm max} = 350$ Myr for $M= 2.5$. These maximal ages are
estimated assuming an optimal initial field strength of $B = 2 \mu$G,
and would decrease for lower or higher field strengths. This estimate
also neglects the increased electron cooling during the compression by
the shock wave, or during earlier stages of the fossil radio plasma
history (as discussed below).  Therefore we expect the radio plasma to
be significantly younger than $t_{\rm max}$.\\ 

\subsection{Possible Origins for the Crosspiece Radio Plasma}

The fact that the weak bow shock wave of 3C 129 could only have
revived relatively young fossil radio plasma gives us a good chance to
identify its origin, because the corresponding radio galaxy could
still be active. There are two candidate sources in our 90cm radio
map: 3C 129.1 and object C.

3C 129.1 resides at the cluster center. Its radio lobes should be
highly buoyant in the cluster atmosphere, similar to those of M87
\citep{churazov01}. Such buoyant bubbles of former activity from 3C
129.1 are a possible explanation for sources A and B in our 327 MHz
map. The faintness of these sources can be naturally understood as the
effect of adiabatic expansion during the buoyant rise in the cluster
atmosphere; using the derived spectral indices $\alpha \approx 2.4$,
the synchrotron luminosity decreases by a factor $D^{-(2+4\alpha)/3}
\approx D^{-3.87}$ due to volume expansion by a factor $D$.

The time required for radio lobe emission to travel from 3C 129.1 to
the Crosspiece's position is $310\, {\rm Myr}/M_{\rm bubble}$, where
$M_{\rm bubble}<1$ is the Mach number of the buoyant bubble's
rise. This is long compared with the maximal fossil radio plasma age,
and probably too long to allow 3C 129.1 to be the Crosspiece's origin.
Additionally, in such a scenario the maximal age of the radio plasma
as calculated should also be significantly corrected downwards due to
increased radiative cooling and adiabatic losses during the early
(more compressed) stage of the rising bubble (see
\cite{2001A&A...366...26E} for a formalism for detailed calculations
of the spectra of aging and expanding radio plasma).

Source C as the origin of the Crosspiece's radio plasma would not have
this difficulty, because it could have passed the present position of
the Crosspiece relatively recently.  The source itself is asymmetric
with an elongation pointing directly towards the Crosspiece. Although
it has a somewhat steeper than normal spectrum for an AGN ($\alpha
\approx 1.5$), it is coincident with a known galaxy.  Therefore we
consider source C a more likely origin of the radio plasma revived by
3C 129's bow shock.

\section{Conclusion}

We present a 330 MHz ($\lambda \sim 90$ cm) radio map of 3C 129 and
its companion 3C 129.1.  In addition to the spectacular structure of
this long NAT source, we also note a small perpendicular object near
the head of the galaxy, which we call the Crosspiece.  This object has
a steep spectrum, and may be a fossil radio plasma revived by
3C 129's bow shock.

This is only one possible explanation for this feature, but it makes a
couple of predictions which can be tested by future observations.
First, the compression of a fossil radio plasma should align the
magnetic fields with the shock surface.  Because this surface is
roughly perpendicular to the line of sight at its front edges, a
relatively clear polarization signature should be visible (see
\cite{eb2001} for numerical simulations of the radio polarization of
shocked radio plasma). Second, if source C is the origin of the
Crosspiece's radio plasma, there may be a low frequency trail
connecting the two. During shock compression, the maximum frequency is
shifted by only $C^{4/3} = 2.2 ... 6.4$ (for $M = 1.4 ...2.3$, see the
change of $\nu$ with $C$ in Eq. \ref{eq:age} for $t_{\rm max}$ fixed),
so a low surface brightness radio emission at and below 100 MHz is
expected. Detailed radio spectra of the pre- and post-shock radio
plasma would give an independent measurement of the shock compression
and geometry in this case.

We note that if this scenario passes the proposed tests, it would help
to disentangle the geometry of the infall of 3C 129 into the X-ray
cluster, because the 3-dimensional velocity can be estimated from the
structure of the Mach cone and the galaxies' redshifts. This would be
an important step forward in our understanding of the peculiar
morphology of 3C 129 in specific, and the nature of radio plasma in
general.

\acknowledgements

WML is a National Research Council Postdoctoral Fellow.  Basic
research in astronomy at the Naval Research Laboratory is funded by
the Office of Naval Research.  DEH acknowledges support from NASA
grant GO1-2135A.  This research has made use of the NASA/IPAC
Extragalactic Database (NED) which is operated by the Jet Propulsion
Laboratory, California Institute of Technology, under contract with
the National Aeronautics and Space Administration.

\null\clearpage

\onecolumn

\null \clearpage

\null

\end{document}